\documentclass{iopart}
\usepackage{graphicx}
\usepackage{caption}
\expandafter\let\csname equation*\endcsname\relax
\expandafter\let\csname endequation*\endcsname\relax	
\usepackage{amsmath}
\usepackage{amssymb}
\usepackage{mathtools}
\usepackage{xcolor}
\usepackage{hyperref}

\renewcommand{\vec}[1]{\ensuremath{\boldsymbol{#1}}} 
\renewcommand{\d}[1]{\, \mathrm{d}#1} 
\newcommand{\pdd}[2]{\frac{\partial #1}{\partial #2}}

\newcommand{\op}[1]{\ensuremath{\mathbb{#1}}}
\newcommand{\smat}{\left( \begin{matrix}}
\newcommand{\emat}{\end{matrix} \right)}

\begin{document}

\title{Measurements of the Solid-body Rotation of Anisotropic Particles in 3D Turbulence}

\author{Guy G. Marcus, Shima Parsa, Stefan Kramel, Rui Ni, Greg A. Voth }  
\address{Department of Physics, Wesleyan University, 
Middletown, Connecticut 06459, USA}
  
\date{\today}

\begin{abstract}
We introduce a new method to measure Lagrangian vorticity and the rotational dynamics of anisotropic particles in a turbulent fluid flow. 
We use 3D printing technology to fabricate crosses (two perpendicular rods) and jacks (three mutually perpendicular rods). 
Time-resolved measurements of their orientation and solid-body rotation rate are obtained from stereoscopic video images of their motion in a turbulent flow between oscillating grids with $R_\lambda$=$91$. 
The advected particles have a largest dimension of 6 times the Kolmogorov length, making them a good approximation to anisotropic tracer particles.
Crosses rotate like disks and jacks rotate like spheres, so these measurements, combined with previous measurements of tracer rods, allow experimental study of ellipsoids across the full range of aspect ratios.
The measured mean square tumbling rate, $\langle \dot{p}_i \dot{p}_i \rangle$, confirms previous direct numerical simulations that indicate that disks tumble much more rapidly than rods.
Measurements of the alignment of crosses with the direction of the solid-body rotation rate vector provide the first direct observation of the alignment of anisotropic particles by the velocity gradients of the flow.
\end{abstract}


\maketitle

\section{Introduction}

The motion of particulate matter in fluid flows has long been a central problem in both fundamental and applied fluid mechanics.  
The dynamics of spherical particles was a natural starting point for Stokes, who considered this problem in the 19th century~\cite{Stokes1851}. A long list of researchers continued work to describe the motion of spherical particles at low Reynolds number, leading to the Maxey-Riley-Gatignol equations~\cite{Maxey-Riley1983,Gatignol1983}.    
The development of numerical simulations~\cite{Yeung1989, Squires1990} and experimental tools~\cite{Dracos1996, LaPorta2001, Mordant2001} capable of revealing the motion of spherical particles in complex flows has led to a recent resurgence of work on particle dynamics~\cite{Toschi2009}.

Extending the problem to non-spherical particles in general is important for many applications ranging from icy clouds~\cite{Pinsky1998}, to bio-locomotion ~\cite{Guasto2012}, to suspension flows in industrial settings~\cite{Lundell2011}.
In 1922, Jeffery analyzed the motion of axisymmetric ellipsoidal particles in Stokes flow~\cite{Jeffery1922}. Their tumbling rate can be expressed as:
\begin{equation}
\dot{p}_i = \Omega_{ij} p_j + \frac{\alpha^2-1}{\alpha^2+1} [S_{ij}p_j - p_i p_j S_{jk} p_k],
\label{eqn:jeffery}
\end{equation}
where $\mathbf{p}$ is a unit vector along the symmetry axis of the ellipsoid, $\alpha$ is the aspect ratio, $\mathbf{\Omega}$ is the anti-symmetric part of the velocity gradient tensor (vorticity), and $\mathbf{S}$ is the symmetric part of the velocity gradient tensor (strain rate).     
A wide range of studies has explored deviations from Jeffery dynamics due to particle inertia, particle shape, and Reynolds number, among other factors~\cite{Leal1980}.	
Analytic work~\cite{Szeri1992,Gustavsson2014} and numerical simulations~\cite{Shin2005,Marchioli2010,Pumir2011,Chevillard2013} have made significant progress extending studies of non-spherical particles in complex flows,  but experimental measurements of the dynamics of anisotropic particles have lagged far behind due to the difficulty of measuring their time-dependent orientation in three dimensions.   
Recently, methods have been developed for measuring time-resolved orientation and position of thin rods in 3D turbulence with stereoscopic optical imaging~\cite{Parsa2012}.  
Another technique uses large transparent anisotropic particles with tracer particles inside and measures their rotations with particle image velocimetry~\cite{Bellani2012}. 

In this paper, we introduce a new way to measure the orientations and rotational dynamics of anisotropic particles that behave like ellipsoids.  
Bretherton  has shown that many anisotropic particles have equivalent ellipsoids so that their tumbling rate follows \autoref{eqn:jeffery} with an effective aspect ratio~\cite{Bretherton1962a}.   
We have identified particles of this class whose orientations can also be directly measured from stereoscopic imaging.    
Three-dimensional printing allows us to create anisotropic particles made of mutually perpendicular thin rods.  
Two perpendicular thin rods form a cross, while three perpendicular rods form a jack. 
Arguments in Bretherton~\cite{Bretherton1962a} and resistive force theory calculations in Appendix A show that a cross rotates like a disk, an ellipsoid with aspect ratio $\alpha  \ll 1$.  
Similarly, a jack rotates like a sphere, which is an ellipsoid with $\alpha=1$.  
We have used stereoscopic video imaging to directly measure the orientation of these objects as a function of time. 
The methods developed in this paper thus allow us to obtain Lagrangian measurements of the full solid-body rotation rate. 

Since a jack rotates like a sphere, its solid-body rotation rate couples only to the vorticity. 
Thus, measurements of the rotation of neutrally buoyant Kolmogorov-scale jacks are direct Lagrangian vorticity measurements---a long sought goal of fundamental fluid mechanics.   
Several other methods for measuring Lagrangian vorticity have also been developed.   
For example, the vorticity optical probe measures the reflections from planar mirrors embedded in spherical particles~\cite{Frish1981}.    
Stereoscopic imaging has been used to measure the rotation of large spheres by applying patterns to the sphere surface~\cite{Zimmermann2011a, Zimmermann2011b} or  embedding small fluorescent tracers in transparent particles~\cite{Klein2013}. 
Our technique of measuring the solid-body rotation of small jacks allows us to accurately measure the fluid vorticity using straightforward imaging methods. 

\section{Experiment}
\label{sec:experiment}

\subsection{Printing 3D Particles}
\label{sec:particles}

We use 3D printing to fabricate anisotropic particles in the shapes of crosses and jacks in order to measure the dynamics of axisymmetric ellipsoids across the range of aspect ratios. 
To print at both high resolution and in high quantity, we used a Connex 500 to make 10,000 of each particle shape. 
Arm lengths were 3 mm, which is 6 times the Kolmogorov length scale in our flow. 
The diameter of any given arm on a rod, cross, or jack is 300$\mu$m, the smallest we could achieve while maintaining the structural integrity of the particles.

To print particles with such small cylindrical arms, it was important to ensure that none of the arms were along the build axes. 
Arms that lay along the vertical build axis had defects and would often break off, while arms lying in the horizontal plane tended to flatten. 
Another important difficulty in printing $\mathcal{O}(10^4)$ particles is removing the support material.  
Connex printers use a different material for the support structure and the printed particles.  The support material can be partially dissolved using a strong base solution (e.g., NaOH) without affecting the particles themselves. 
We found that using an ultrasonic bath made the removal process much more efficient, and the particles could be filtered out of the solution with almost no loss.  

In order to have neutrally buoyant particles, the density of our fluid is matched to the particle density. 
The density is adjusted by adding calcium chloride to water~\cite{Parsa2012}.
We used the print material VeroClear, whose bulk density was quoted at 1.17~g/cm$^3$. 
However, we found that the manufacturer's quote differed significantly from the density of the fluid in which the particles were neutrally buoyant.  
After the particles were immersed in the fluid for several hours, we found that different particles were neutrally buoyant at slightly different fluid densities ranging from 1.21 to 1.23~g/cm$^3$. 
We chose the density of the fluid for the experiment as 1.22~g/cm$^3$ based on the population average.
More work is needed to understand the mass density distribution inside 3D printed objects, but our particles are sufficiently density matched that their rotations should accurately represent the neutrally buoyant case~\cite{Voth2002}.    
To make the particles fluorescent, they were placed in a high concentration Rhodamine solution at elevated temperature (60-80$^\circ C$) for several hours. 

\subsection{Turbulent flow between oscillating grids}
\label{sec:flow}

\begin{figure}
\centering
\includegraphics{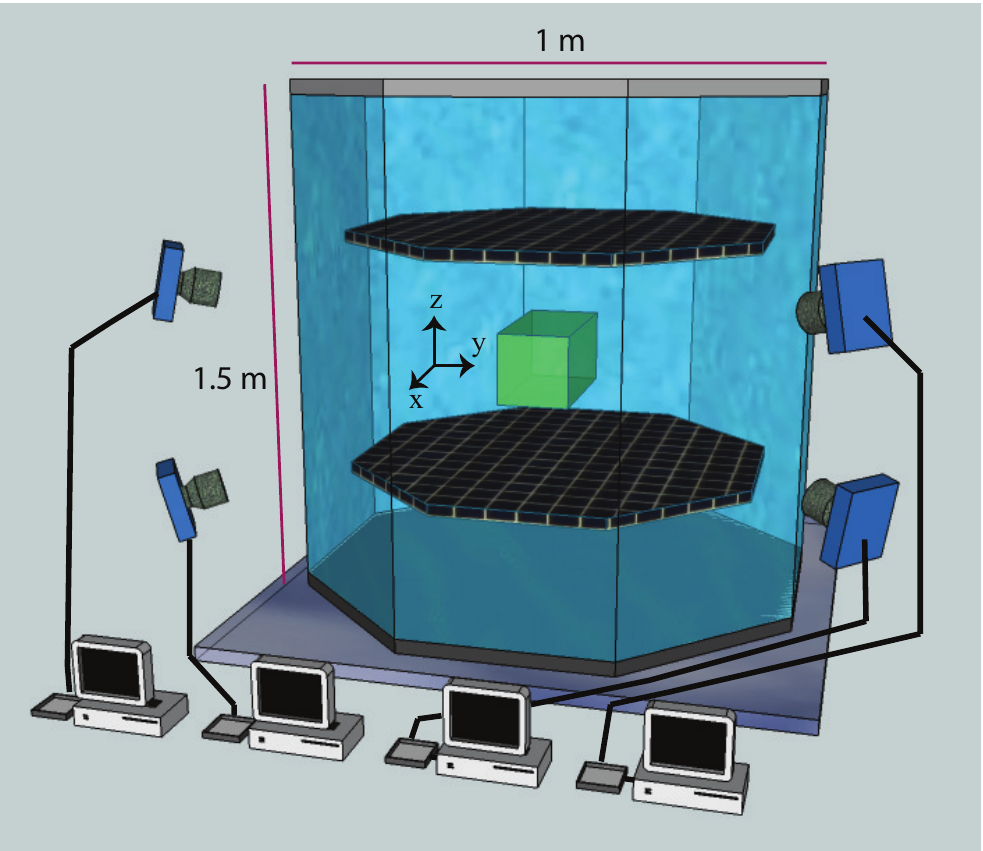}
\caption{Experimental setup (figure from \cite{Wijesinghe2012}). In the octagonal flow between oscillating grids, a central viewing volume in the focus of the four stereoscopically arranged video cameras is illuminated by a green Nd:YAG laser.}
\label{fig:tank}
\end{figure}

The turbulence is generated in a $1\times 1 \times 1.5$ m$^3$ octagonal prism using two parallel 8\,cm-mesh grids oscillating in phase (see  \autoref{fig:tank}). 
The measurements presented here were performed at a grid frequency of 1 Hz, which produced a flow with a root mean squared (rms) velocity $U \equiv  \sqrt{\langle u_i u_i \rangle} /3=  20 $ mm/s and energy injection length scale, $L=U^3/\varepsilon=60$ mm so that  $R_\lambda \equiv (15 U L/\nu)^{1/2} = 91$.  
The energy dissipation rate, $\varepsilon$ = 133 mm$^2$s$^{-3}$, was measured from the mean square tumbling rate of jacks as described in \autoref{sec:results}. 
The kinematic viscosity of the CaCl$_2$ solution that is approximately density matched to the particles is $\nu = 2.17$ mm$^2$/s.
The Kolmogorov length scale is $\eta = (\nu^3 / \varepsilon)^{1/4} = 526~\mu$m, and the Kolmogorov time scale is $\tau_\eta = (\nu / \varepsilon)^{1/2} = 128$~ms. 
We chose this low Reynolds number to ensure that our particles were only 6$\eta$, where the mean square rotation shows only a small deviation from the tracer limit ~\cite{Shin2005,Parsa2014}.

We use four stereoscopic cameras (1280$\times$1024 at 450 Hz) with a custom real-time image compression system~\cite{Chan2007, Wijesinghe2012} to allow continuous imaging.
A green Nd:YAG laser with 50 W average power illuminates a detection volume of roughly $3 \times 3 \times 3$~cm$^3$ at the center of the tank, where the flow is quite homogeneous~\cite{Blum2010a}.    
We used two perpendicular expanded beams in the horizontal plane, each reflected back upon itself, in order to provide 4 directions of illumination and minimize shadowing of some arms of a particle by other arms.     
With an average particle number density of only $5\times 10^{-3}$~cm$^{-3}$, there was a particle in view less than 20\% of the time, which made the image compression system essential for acquiring enough trajectories to converge statistics~\cite{Chan2007}.   

\subsection{Image Analysis.}
\label{sec:analysis}

\begin{figure}[h]
\centering
\includegraphics[width=0.8\textwidth]{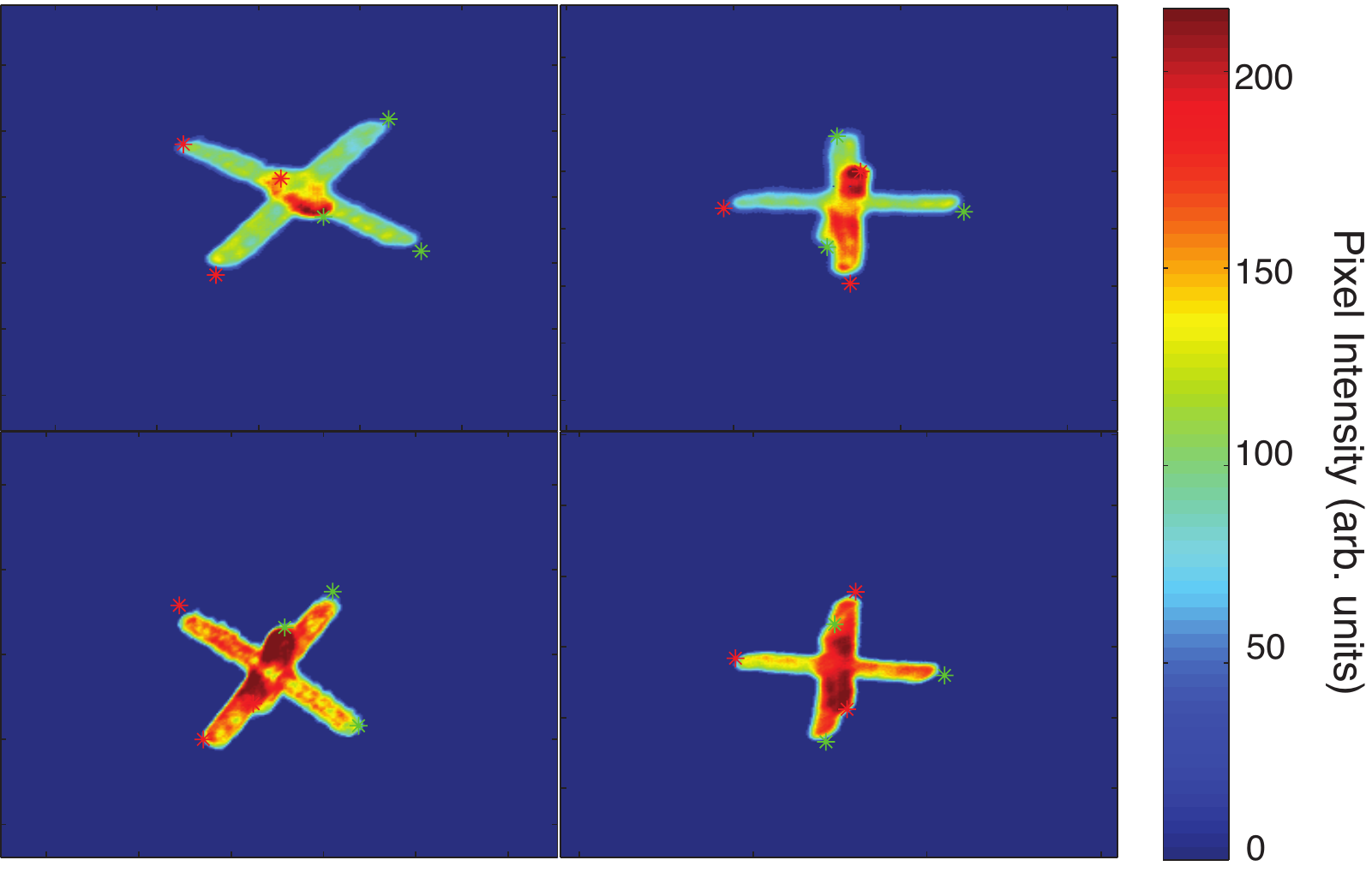} 
 \caption{A sample image of a jack from each of the four cameras. The pairs of red and green asterisks denote the two ends of an arm determined by the orientation-finding algorithm.}
\label{fig:projex}
\end{figure}    

\emph{Orientation.}---An example of a jack imaged by all four cameras is shown in \autoref{fig:projex}.  
When a particle is visible on all four cameras, we find the three dimensional position of the particle using stereomatching methods~\cite{Ouellette2006}.  
We developed a nonlinear fitting algorithm to determine the orientation of a particle from a set of stereoscopic images. 
Any orientation is specified by a rotation matrix, $\op{O}$, which can be parameterized by three Euler angles $(\phi, \theta, \psi)$~\cite{Goldstein}.  
From a measurement of the particle's center and an initial guess of its orientation, we construct a model of the particle and project it onto each of the four image planes using the calibrated camera parameters~\cite{Tsai1987}. 
The total difference in intensity between a model image and an experimental image provides the residual that is minimized by a nonlinear search in Euler angle space. 

To minimize computational cost, we project the endpoints of each arm onto the image plane of each camera, and then model the intensity distributions in two dimensions.  
The model image of an arm is formed by a Gaussian intensity distribution along the width of each arm (in 2D) and a Fermi intensity distribution across the length of the arm. 
For jacks and crosses, each arm in the model has an identical intensity distribution.
As can be seen from \autoref{fig:projex}, the arms in the experimental images are not of uniform intensity.
The observed intensity has a non-trivial dependence on the angles between the arms, the illumination, and the viewing direction~\cite{Parsa2014NPr}, which has not yet been included in our model. However, we found the simple model adequate enough for our purposes.  

The orientation-finding algorithm must be seeded with an initial guess for the orientation.   
Except for the first frame of a track, we use the previous frame as the initial guess, since the rotation between frames is less than 0.01 radians.
The initial guess for the beginning of a particle track is reliably generated using an optical tomographic reconstruction algorithm~\cite{Elsinga2006}.  
We also compared tomographic reconstruction as a method for finding particle orientation, but we found it to be both less accurate and more computationally expensive than our own algorithm.   

\begin{figure}
\centering
\includegraphics[width=\textwidth]{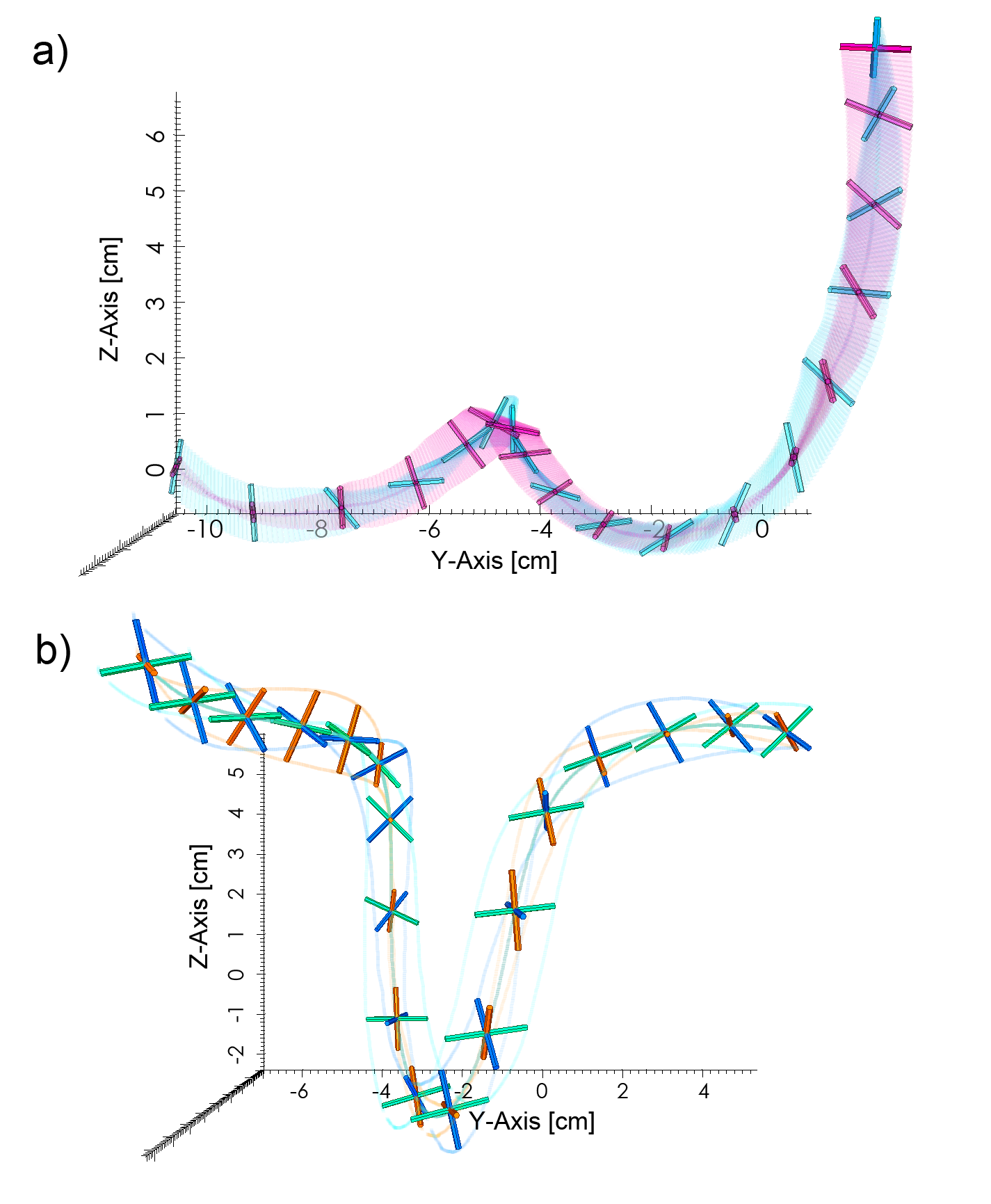}
\caption{\textbf{(a)} A reconstructed trajectory of a cross in three-dimensional turbulence. 
The two different color sheets trace out the path of the particle through space and time. 
The length of the particle track is 336 frames, or $5.7 \,\tau_\eta$. A cross is shown every 15 frames. 
\textbf{(b)} A reconstructed trajectory of a jack in three dimensional turbulence. 
The three different colors distinguish the arms of the jack and trace out their path as the particle rotates. 
The dark green line denotes the trajectory of the jack's center. 
The length of the particle track is 1025 frames, or $17.5 \,\tau_\eta$. 
A jack is shown every 50 frames. (\emph{Note:} neither the crosses nor the jacks shown above are drawn to scale.)}
\label{fig:cross+jack_track}
\end{figure}

The Euler angles found for a jack give 1 of the 24 orientations related by symmetry.    
To see this, consider 1 of the 6 arms of a jack and define it by the vector ${\vec{p}}=\hat{\vec{z}}$; there are 4 symmetrically identical orientations obtained by rotations of $\pi/2$ about the $z$ axis. 
There are 4 such orientations for each of the 6 arms, for a total for 24.
A cross has 8 symmetrically identical orientations, and a rod has 2.   
We ensure that we have a consistent series of orientations along a single trajectory by comparing the orientation found in each frame with that from the previous frame, and choose the orientation of the particle that produces the smallest total rotation between frames.   

\autoref{fig:cross+jack_track}(a) shows a representative track for a cross, which is $5.7\,\tau_\eta$ long. 
It demonstrates the effectiveness of our algorithm at determining the full range of orientations.
At various points along the track, the rotation of the cross about a vector nearly coplanar with its two arms is clearly visible. 
In \autoref{fig:cross+jack_track}(b) we also show an example jack trajectory that is $17.5\,\tau_\eta$ long.

\emph{Solid-body rotation rate.}---Once we have measurements of orientation and position at each time step, the natural quantity to consider  is the rate of change of the unit vector defining the particle's orientation, which we call the tumbling rate, $\dot{\vec{p}}$. 
For crosses and jacks, we can also measure the full solid-body rotation rate vector, $\vec{\omega}_s$, from a series of orientation measurements smoothed along the particle's trajectory.
One method for doing this has been described in~\cite{Zimmermann2011b}.   
We take a different approach using the tools we already developed for least squares optimization in Euler angle space.  
The  problem can be framed as finding the initial orientation matrix, $\op{O}(t_i)$ and the rotation matrix over a single time step, $\mathbb{R}$,  that together give the particle orientation matrix as a function of time, 
\begin{align}
\op{O}(t) =   \mathbb{R}^{\frac{t-t_i}{\tau_f}}\op{O}(t_i),
\end{align}
where $\tau_f$ is the period between images. 
A non-linear least squares fit is used to find the Euler angles for the matrices $\op{O}(t_i)$ and $\mathbb{R}$ that best match the measured orientation matrices.   
Then, the rotation matrix $\mathbb{R}$ can be decomposed as a rotation by an angle $\Phi$ about the solid body rotation axis, $\hat{\vec{\omega}}_s$, in accordance with Euler's theorem~\cite{Goldstein}, from which we obtain the magnitude of the solid body rotation rate, $\omega_s = \Phi / \tau_f$.  

\begin{figure}[t]
\centering
\includegraphics[width=\textwidth]{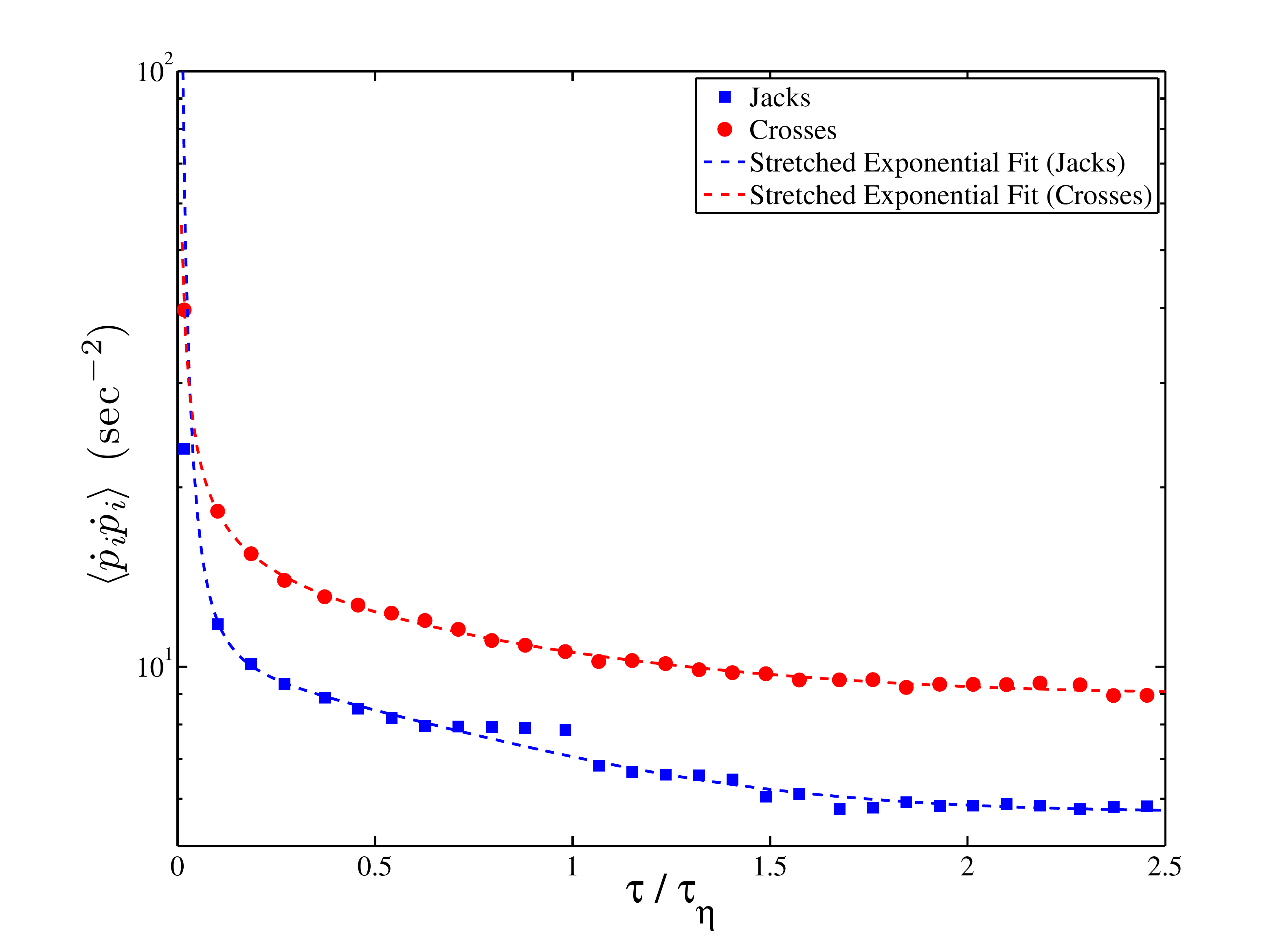}
\caption{Measurements of the mean square tumbling rate, $\langle \dot{p}_i \dot{p}_i \rangle $, as a function of the fit length for both  jacks (blue) and crosses (red). To determine the true value, we extrapolate to zero fit-length by fitting the data to a stretched exponential~\cite{Voth2002}. }
\label{fig:ppfl}
\end{figure}

The solid-body rotation rate, $\vec{\omega}_s$, is related to the tumbling rate by $\dot{\vec{p}}=\vec{\omega}_s \times \vec{p}$. 
The difference between the two quantities is that $\dot{\vec{p}}$ does not depend on the vector component of $\vec{\omega}_s$ lying along $\vec{p}$. 
We use this relationship to determine the tumbling rate from measurements of particle orientation and their solid-body rotation rate. 
In order to correct for the contributions from orientation measurement errors to a measurement of the mean square tumbling rate, one needs to measure $\dot{\vec{p}}$ over a a range of fit-lengths, $\tau$~$=t-t_i$.
\autoref{fig:ppfl} shows our measurement of $\langle \dot{p}_i \dot{p}_i \rangle$ as a function of $\tau$. The solid lines are fits of the function,
\begin{align}
 f(\tau) = A \tau^B + C \exp \left( D \tau + E \tau^2 \right),
\end{align}
where the first term models the random error that dominates at small $\tau$ and the second term models the effect of filtering experimental measurements of orientation~\cite{Voth2002}.
The fit parameter $C$ gives an estimate of the function value at $\tau=0$ if there were no random errors. 

\section{Results}
\label{sec:results}

\begin{figure}[h]
\centering
\includegraphics[width=\textwidth]{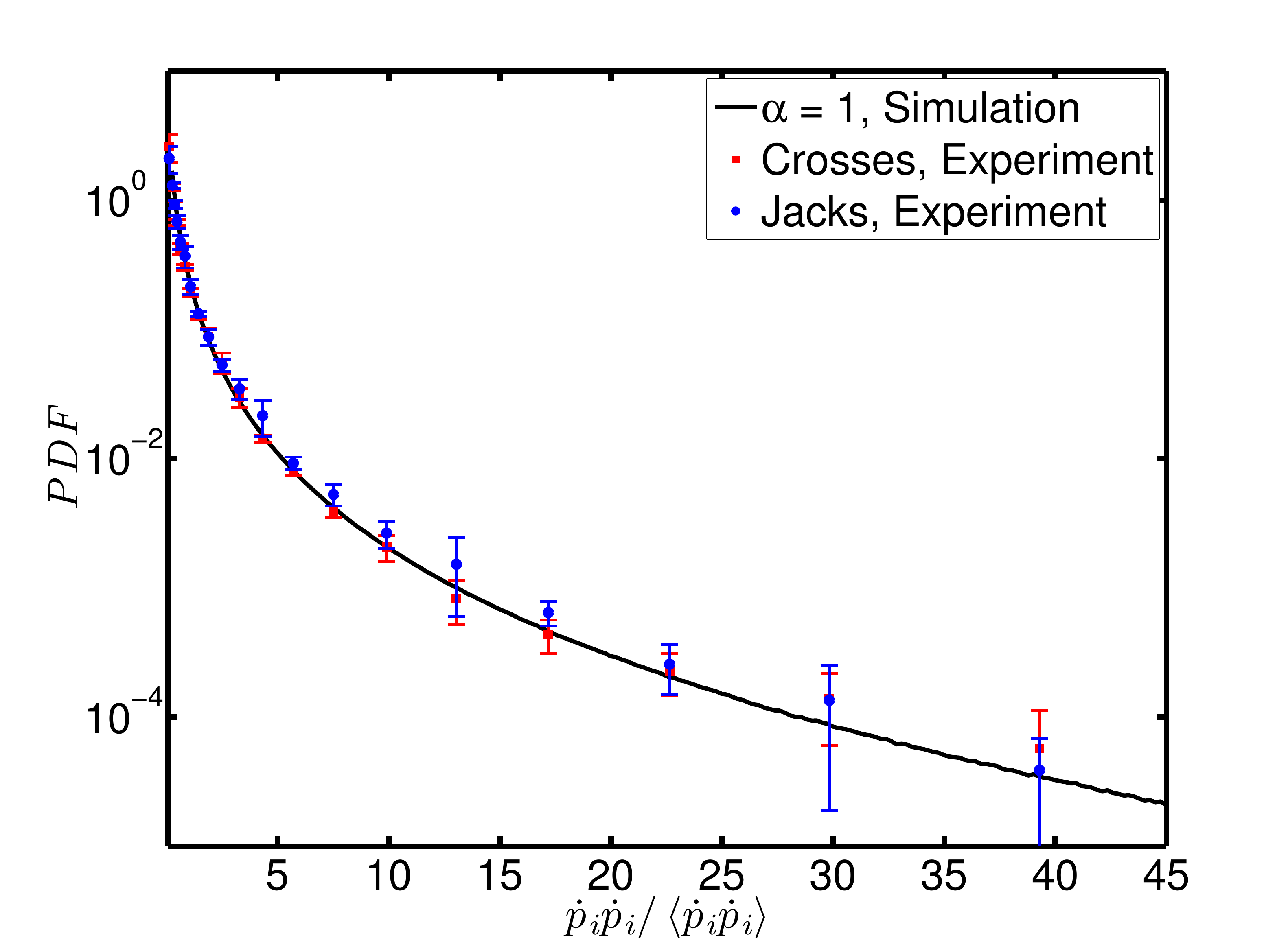}
\caption{The PDF of the mean square tumbling rate for our experimental measurements of crosses (red squares) and jacks (blue circles) as well as direct numerical simulations of spheres (solid line).}
\label{fig:PDF_pd2}
\end{figure}

From the measured solid-body rotation rates of many jacks and crosses, we can obtain the probability density function (PDF) of the squared tumbling rate, which is shown in \autoref{fig:PDF_pd2}. 
Also shown is the PDF obtained from direct numerical simulations of spheres~\cite{Parsa2012}.   All three PDFs agree within experimental uncertainties.  
Numerical work has shown that there should be slightly more probability density in the tail of the PDF for rods and disks than there is for spheres ~\cite{Parsa2012}.  For large tumbling rates, our data shows the opposite, with jacks having slightly higher probability than crosses, although the sources contributing to this small discrepancy are known. 
The error bars shown in \autoref{fig:PDF_pd2} account for random error as well as the systematic error that results from the fit-length dependence of the tumbling rate measurements. 
An additional source of error that is not included in the error bars is the self-shadowing of particles as they pass through certain orientations dependent on the camera configuration. 
The most dramatic cases are when an entire arm is missing from each of the four cameras, such that a jack, for example, will look like a cross along part of a given trajectory.
The reduction in accuracy in determining these particular orientations occasionally leads to erroneously high measurements of the solid-body rotation rate, which pushes additional probability density towards the tail of the PDF. 
This effect is stronger for jacks than for crosses, which is why the jack PDF is slightly higher in our measurements at large tumbling rates. 	

\begin{figure}
\centering
\includegraphics[width=\textwidth]{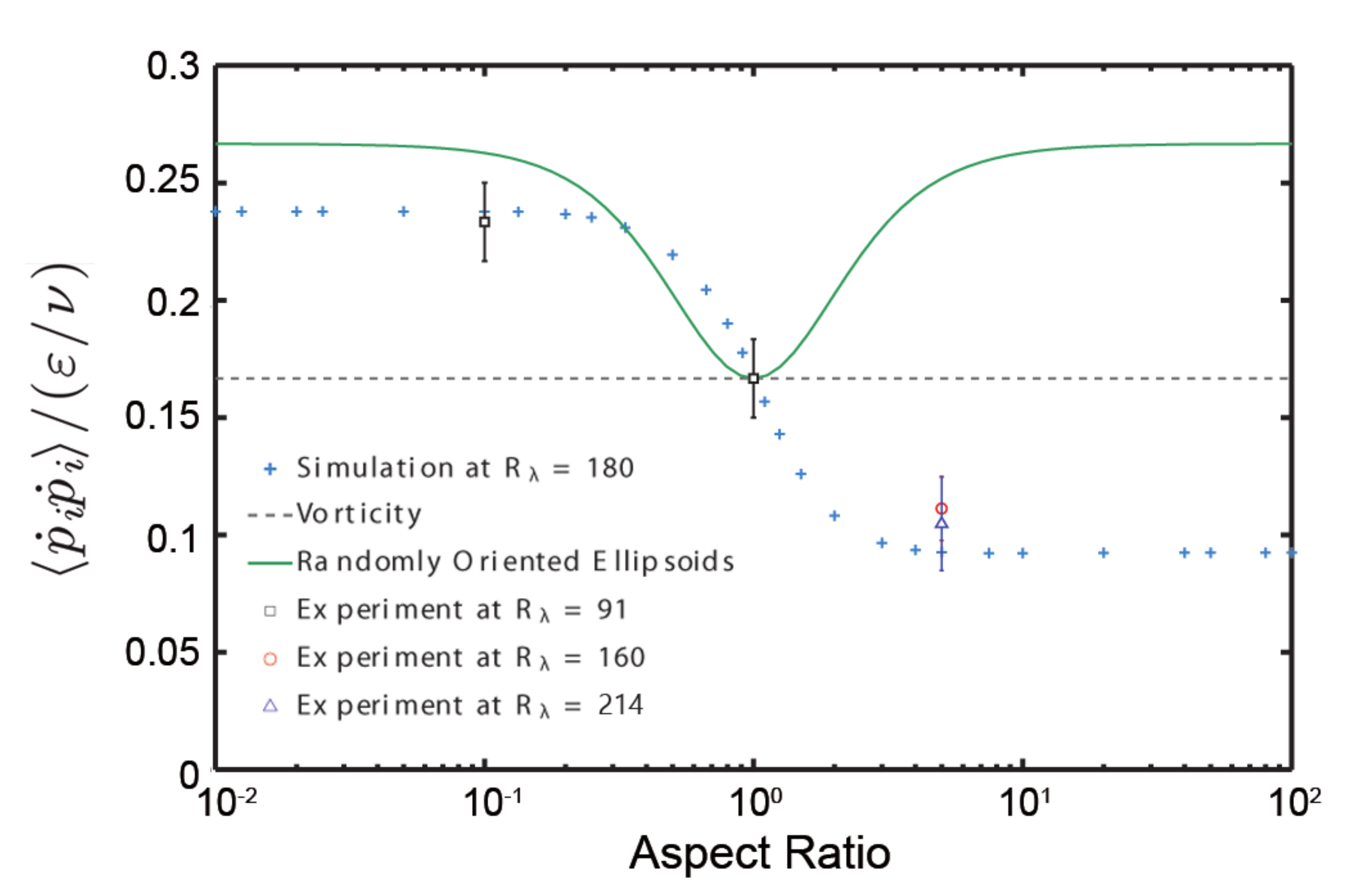}
\caption{Mean square tumbling rate as a function of aspect ratio.  Solid squares are our data for crosses and jacks.  The other data is from \protect{\cite{Parsa2012}}.  Plus symbols are numerical simulations.  The triangle and circle are experimental measurements of rods at $R_\lambda=214$ and 160.  The solid line is an analytic result for randomly oriented ellipsoids. } 
\label{fig:pd2_alpha}
\end{figure}

Previous work by Parsa \emph{et al.} showed that the mean square tumbling rate for axisymmetric ellipsoids, $\left< p_i p_i \right>$, is strongly affected by alignment of anisotropic particles by turbulence~\cite{Parsa2012}.
They found agreement between simulations and experimental measurements of rods, but particles with a wider range of aspect ratios could not be measured experimentally  with the tools available at that time. 
We have now measured the mean square tumbling rate of crosses and jacks using the techniques described in \autoref{sec:experiment}. 
In \autoref{fig:pd2_alpha}, we show the mean square tumbling rate normalized by the Kolmogorov timescale as a function of aspect ratio. 
Our measurements show crosses tumbling at a considerably higher rate than jacks or rods, but still more slowly than the randomly-oriented prediction.   There is good agreement with simulations across the full range of aspect ratios.

In turbulence experiments, it is always a challenge to measure the energy dissipation rate, which appears in the normalization of the vertical axis in  \autoref{fig:pd2_alpha}.  
We attempted to make independent measurements using non-fluorescent tracer particles, but did not succeed because of reduced light scattering from tracers in the density-matched CaCl$_2$ solution.   
However, the measurements of  jack rotations provide a new way to measure the energy dissipation rate. 
Because they rotate like spheres, they give a direct measurement of the vorticity, and in isotropic turbulence their vorticity is directly related to the energy dissipation through $\langle \Omega_{ij} \Omega_{ij} \rangle$=$\langle S_{ij}S_{ij} \rangle=\varepsilon/\nu$. This implies that for spheres $\langle \dot{p}_i \dot{p}_i \rangle = \frac{1}{6} \varepsilon/\nu$~\cite{Parsa2012}.
We use this to determine our energy dissipation rate.
This makes our jack data at $\alpha=1$ match the simulations by definition, and the agreement of the cross data at aspect ratio $\alpha=0.1$ with numerical simulations is an independent result of the measurement.

The solid green curve in \autoref{fig:pd2_alpha} gives the mean square tumbling rate as a function of aspect ratio for randomly oriented ellipsoids. 
While the tumbling rate of jacks is unmodified from the randomly oriented case, both rods and crosses show smaller tumbling rates due to the effects of alignment by turbulence.   
Shin and Koch were the first to notice that the tumbling rate of rods that have correlated with a turbulent flow is reduced in comparison to that of randomly oriented rods~\cite{Shin2005}. 
More recent numerical work showed that the effects of alignment persist across the full range of aspect ratios~\cite{Parsa2012}.
The leading-order effects are contained in the Lagrangian three-point correlations of the velocity gradients, which imply higher tumbling rates for disks than for rods~\cite{Gustavsson2014}.
A number of studies on the dynamics of rods show that they align with the vorticity, and this has been used to explain the slower tumbling rate of rods since the component of the vorticity along the rod axis does not contribute to its tumbling rate~\cite{Pumir2011, Chevillard2013, Wilkinson2012, Ni2014}.   
We find that alignment with vorticity is also responsible for the reduction of the tumbling rates for crosses. 

\begin{figure}[t]
\centering
\includegraphics[width=\textwidth]{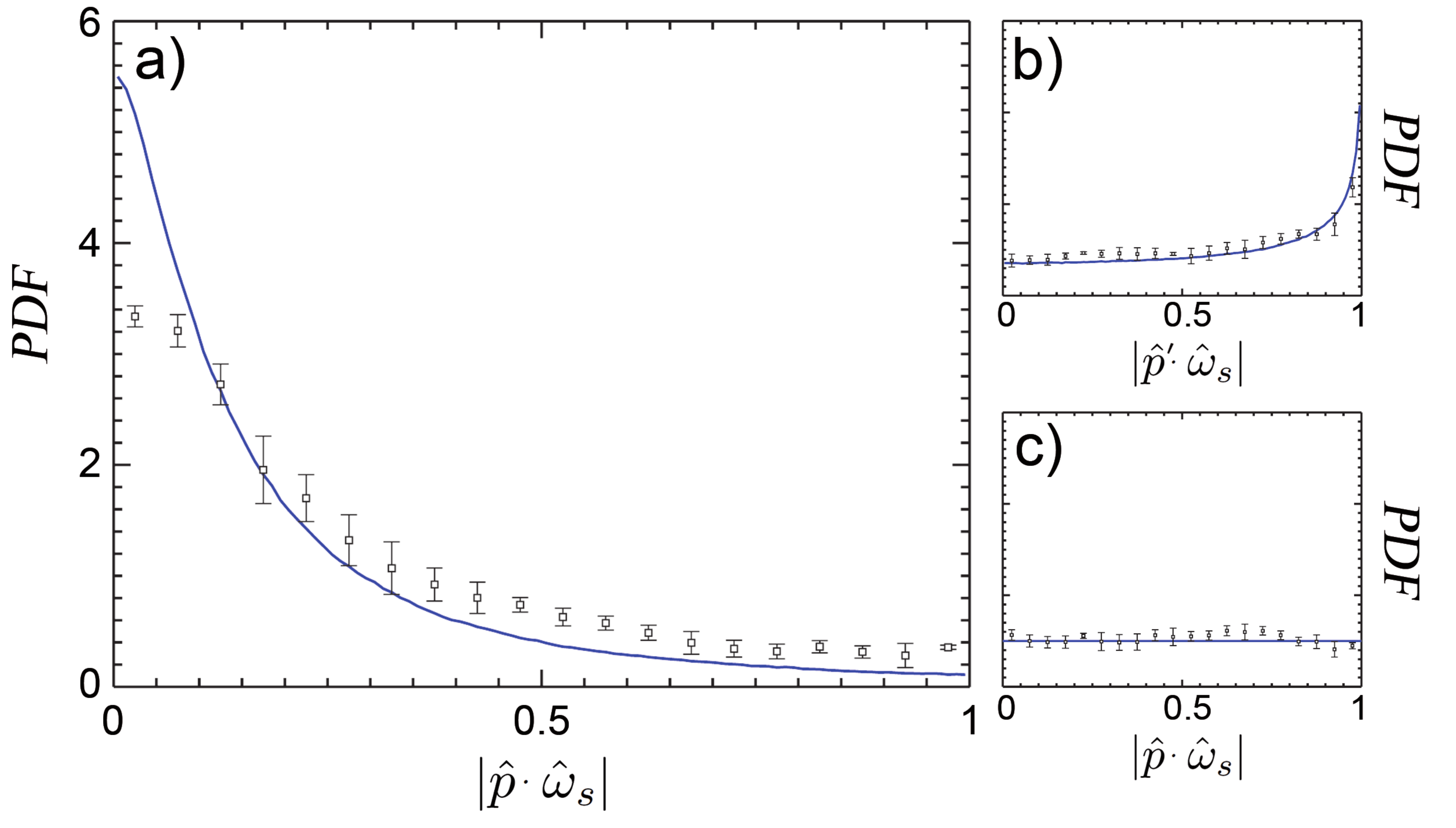}
\caption{The PDF of the alignment between a particle's orientation and its solid body rotation rate, $|\hat{\vec{p}} \cdot\hat{\vec{\omega}}_s|$, for (a) crosses and (c) jacks. Symbols are experimental measurements and solid lines are numerical simulations.  (b) shows the alignment of one of the arms of a cross, $\vec{p}^\prime$, with its solid body rotation rate vector. }
\label{fig:cospw_all}
\end{figure}

We can directly measure the preferential alignment of a particle by measuring the angle between the orientation of the particle and its solid body rotation rate vector.
In \autoref{fig:cospw_all}, we plot the PDF of the magnitude of the cosine of this angle,  $\lvert \vec{p} \cdot \hat{\vec{\omega}}_s \rvert$,  for both crosses and jacks, and compare them with numerical simulations. 
The peak near $\lvert \vec{p} \cdot \hat{\vec{\omega}}_s \rvert=0$ for crosses in \autoref{fig:cospw_all}(a) shows that disks preferentially align with $\vec{p}$ perpendicular to $\vec{\omega}_s$.   
\autoref{fig:cospw_all}(b) confirms the same story as \autoref{fig:cospw_all}(a).  
It shows that an arm of a cross is preferentially aligned with the solid body rotation rate vector.  
The  height of the peak in \autoref{fig:cospw_all}(b) is lower than that in \autoref{fig:cospw_all}(a) because any vector in the plane of the disk is equally likely to align with $\vec{\omega}_s$.   
In a turbulent flow, disks rotate like a spun coin rather than a tossed frisbee. 
For jacks in \autoref{fig:cospw_all}(c), there is no preferential alignment because they rotate like spheres.   Numerics and experiment are in quite good agreement.
The deviations near $\lvert \vec{p} \cdot \hat{\vec{\omega}}_s \rvert=0$ for crosses are likely the result of measurement error in the solid body rotation rate vector.

The observed alignment in \autoref{fig:cospw_all} seems natural if it is thought of as a result of the Lagrangian stretching of the fluid.  Stretching will align both the vorticity and a long axis of a particle with the stretching direction~\cite{Ni2014}.   For rods this means that $\vec{p}$ aligns with the vorticity, which decreases the tumbling rate.  
Because disks have their long axis perpendicular to $\vec{p}$, they preferentially have $\vec{p}$ perpendicular to the vorticity, which makes $\vec{p}$ also preferentially perpendicular to the solid-body rotation rate.   This creates a larger tumbling rate for disks since $\vec{\dot{p}}=\vec{\omega}_s \times \vec{p}$, consistent with the data in \autoref{fig:pd2_alpha}.   

Although disks tumble faster than rods, they still  tumble slightly slower than if they were randomly oriented, as seen in~\autoref{fig:pd2_alpha}. 
Since disks have a much smaller difference from the randomly oriented case, one might conclude that disks are less strongly aligned than rods by the turbulence.  
However, this is largely a result of the way $\dot{\vec{p}}$ is defined.   
We have measured the normalized mean square tumbling rate of a unit vector along one of the arms of a cross as $0.12 \pm 0.02$, which is smaller than the tumbling of either crosses or jacks and the same as measurements for rods within experimental error.  
This explicitly indicates that crosses (disks) are also strongly aligned by turbulence, and it is only the definition of $\vec{p}$ as perpendicular to the two arms that increases the mean square tumbling rate so much in comparison to rods.  
It is not easy to directly compare the degree of alignment of rods and disks because disks have an entire plane that can align with the stretching, while rods have only a director.
The picture that emerges from our data and previous work is that anisotropic particles are aligning with the Lagrangian stretching direction, and this suppresses the tumbling rate for all particles except nearly spherical oblate ellipsoids.  
The amount of suppression is strongly dependent on the axis whose tumbling rate is considered.

\section{Conclusions}
We have developed a method for measuring the time-resolved Lagrangian orientation and solid body rotation rate of anisotropic particles in a turbulent flow.  
By measuring the rotation of 3D printed jacks and crosses we are able to extend previous measurements of rods to cover the full range of aspect ratios of axisymmetric ellipsoids.  
Moreover, we have provided a way to directly probe Lagrangian vorticity with a single particle measurement, which has potential for application in a wide range of flows and Reynolds numbers.   
We find that the mean square tumbling rate, $\langle \dot{p}_i \dot{p}_i \rangle$, agrees with DNS data at points spanning the full range of aspect ratios.
Our measurements show that crosses are preferentially aligned in turbulence with their orientation axis perpendicular to their solid body rotation rate vector.   
To the best of our knowledge, this is the first direct experimental measurement of the preferential alignment of particles with their rotational motion in a turbulent flow.  
Our results support a natural picture of alignment in turbulence where particles have their long axes aligned with the Lagrangian stretching direction of the fluid flow.      

\appendix{

\section{Resistive Force Theory}
\label{sec:res-force}

Resistive force theory is widely used to model the motion of slender bodies at small Reynolds number (see, \emph{e.g.},~\cite{Johnson1979}).
We will apply it first to a small thin rod of width $a$ in a general velocity field.  
Then we will generalize to multiple perpendicular rods in order to show that jacks rotate like spheres and crosses rotate like disks. 

Resistive force theory assumes that the differential viscous drag forces acting on a small segment of a slender body are ~\cite{Gray1955}:
\begin{align}
\d{\vec{F}} &= 2aC_d \vec{u_\perp} \d{r} \nonumber \\
\d{\vec{f}} &= aC_d \vec{u_\parallel} \d{r}
\label{eqn:res-force}
\end{align} 
where $\d{\vec{f}}$ is parallel to the rod, while $\d{\vec{F}}$ is perpendicular to the rod and responsible for producing all of the torque. 
The vector \vec{u} is the relative velocity between the segment and the fluid, and its subscripts indicate the component parallel or perpendicular to the slender object. 

We can use these expressions to calculate the total torque due to fluid flow past the rod. 
Since the particles are on the order of the Kolmogorov scale, we linearize the velocity field:
\begin{equation}
{v}_i(\vec{x}) = {v}_i(\vec{x_\circ}) + \pdd{v_i}{x_j} \bigg|_{\vec{x}=\vec{x_\circ}} \Delta{x}_j \nonumber.
\end{equation}
Taking the center of the particle to be at rest at the origin, the velocity can be written as:
\begin{equation}
{v}_i(\vec{x}) = \pdd{v_i}{x_j} \bigg|_{\vec{x}=0} r{p}_j,
\end{equation}
where $\vec{p}$ is the normalized orientation vector of the rod and $r$ is the radial distance from the center of the rod.    
The drag force is determined by the relative velocity between the fluid and the rigid-body rotation of the rod:
\begin{equation}
\vec{u} = \vec{v} - \vec{\omega}_s \times r \vec{p}.
\end{equation}

The toque comes from the component of velocity perpendicular to the rod, $\vec{u}_\perp$. 
We can write this by subtracting the parallel component from the full relative velocity.  
Entering $\vec{u}_\perp$  into \autoref{eqn:res-force}, we have:
\begin{align}
\d{{F}_i} &= 2aC_d \left[ r \pdd{u_i}{x_j} p_j - \left( r \pdd{u_k}{x_l} p_l p_k \right) p_i - r \epsilon_{imn} (\omega_s)_m p_n  \right] \d{r} \nonumber \\
&= \left[ \left( S_{ij} + \Omega_{ij} \right)p_j - p_k \left( S_{kl} + \Omega_{kl} \right) p_l p_i - \epsilon_{imn} (\omega_s)_m p_n \right] 2aC_d r\d{r},
\label{df1}
\end{align}
where we have written the velocity gradient in terms of its symmetric and antisymmetric parts, $S_{ij}$ and $\Omega_{ij}$, respectively.%

We now rearrange \eqref{df1}, noting that $p_k \Omega_{kl} p_l = 0$ since $p_k p_l$ forms a symmetric tensor while $\Omega_{kl}$ is antisymmetric, and compute the differential torque element:%
\begin{align}
\d{\tau}_q &= \epsilon_{qti} rp_t \d{F}_i \nonumber \\
&= \epsilon_{qti} p_i \left[ \Omega_{ij}p_j + S_{ij}p_j - p_i p_k S_{kl}p_l - \epsilon_{imn} (\omega_s)_m p_n \right] 2aC_d r^2 \d{r}.
\end{align}
Integrating along the length of the rod, $l$, we have the total torque,
\begin{align}
\tau_q &= \epsilon_{qti} p_t \left[ \Omega_{ij}p_j + S_{ij}p_j - p_i p_k S_{kl}p_l - \epsilon_{imn} (\omega_s)_m p_n \right] 2aC_d \int^{l/2}_{-l/2} r^2 \d{r} \nonumber \\
&= \epsilon_{qti} p_t \left[ \Omega_{ij}p_j + S_{ij}p_j - p_i p_k S_{kl}p_l - \epsilon_{imn} (\omega_s)_m p_n \right] \dfrac{a}{6}C_d l^3 
\label{rod_torque_omega}
\end{align}
Now, consider the term $\epsilon_{imn} (\omega_s)_m p_n = \left( \vec{\omega}_s \times \vec{p} \right)_i$. This is nothing else than the rate at which the orientation of the rod is changing, $\dot{\vec{p}}$. Thus, we have:
\begin{equation}
\tau_q =  \epsilon_{qti} p_t \left[ \Omega_{ij}p_j + S_{ij}p_j - p_i p_k S_{kl}p_l - \dot{{p}_i}  \right] \dfrac{a}{6}C_d l^3 
\label{rdtrq}
\end{equation}

For any orientation, $\vec{p}$, \eqref{rdtrq} can be used to compute the $q$th component of the torque on a single rod.  For Stokes flow:
\begin{equation}
\tau_q = 0,\quad \text{for q=1,2, and 3}.
\end{equation}
When we impose this constraint on \eqref{rdtrq}, we immediately recover Jeffery's equation (\autoref{eqn:jeffery}) in the limit of infinite aspect ratio (i.e., for a rod):
\begin{align}
\dot{{p}_i} = \Omega_{ij}p_j + S_{ij}p_j - p_i p_k S_{kl}p_l
\label{rjef}
\end{align}

With this established, we can now explore the rotational dynamics of objects that are composites of perpendicular rods with arbitrary lengths. 
From \eqref{rod_torque_omega}, we can read off the $q$th component of the torque on the $\alpha$th arm, $\tau_q^\alpha$. 
The total $q$th component of the torque is then:
\begin{equation}
\tau_q = \sum\limits_{\alpha=1}^N \Gamma^\alpha \epsilon_{qti} p_t^\alpha \left[ \Omega_{ij}p_j^\alpha + S_{ij}p_j^\alpha - p_i^\alpha p_k^\alpha S_{kl}p_l^\alpha - \epsilon_{imn} (\omega_s)_m p_n^\alpha \right] 
\label{atr}
\end{equation}
where $N \le 3$ is the number of arms on the particle, $l_\alpha$ is the length of the $\alpha$th arm, and $\Gamma^\alpha$ is the pre-factor for the $\alpha$th arm, $\Gamma^\alpha \equiv a C_d l_\alpha^3$.
From here, we can extend the calculation made for single rods to particles composed of $N$ rods. 
The zero net torque condition of Stokes flow still holds true, but there will now be contributions from multiple arms because of the sum over $\alpha$. 
If we consider axisymmetric particles, $l_1 = l_2 \neq l_3$, and are careful to define $\vec{p}$ along the symmetry axis of the particle, \autoref{atr} leads to an analog of Jeffery's equation:
\begin{equation}
\dot{p}_i = \Omega_{ij} p_j + \frac{\alpha_p^3-1}{\alpha_p^3+1} [S_{ij}p_j - p_i p_j S_{jk} p_k],
\end{equation}
where $\alpha_p \equiv l_3 / l_1$. 

We find that resistive force theory for the rotational motion of a set of three perpendicular rods with $l_1 = l_2 \neq l_3$ reproduces Jeffery's equation, except that the aspect ratio appears with cubic dependence rather than quadratic, such that the particles behave like ellipsoids with an effective aspect ratio~\cite{Bretherton1962a}.     
The cases we study are jacks, $l_1 = l_2 = l_3$ with $\alpha_p=1$, and crosses, $l_1 = l_2$ but $l_3=0$ with $\alpha_p=0$ .   
These both happen to be cases where $\alpha_p^3 = \alpha_p^2$, so, within the limits of resistive force theory, we can simply say that jacks rotate like spheres and crosses rotate like disks.

}

\ack

It is our pleasure to thank Jonas Einarsson, Peko Hosoi, Nadia Cheng, Bernhard Mehlig, Vikram Singh, and Federico Toschi for helpful discussions.  
We also thank Federico Toschi and Enrico Calzavarini for providing us with the DNS
data. 
We acknowledge support from the Wesleyan McNair program, US NSF grant DMR-1208990, and COST Actions MP0806 and FP1005.  

\section*{References}
\bibliographystyle{unsrt}
\bibliography{anisotropic.bib}

\end{document}